\documentclass[11pt]{article}
\usepackage{amssymbols}
\usepackage{fullpage}

\makeatletter

\newtheorem{theorem}{Theorem}[section]
\newtheorem{lemma}[theorem]{Lemma}

\newtheorem{corollary}[theorem]{Corollary}

\newtheorem{definition}[theorem]{Definition}

\newcommand{\NP}{\mathbf{NP}}
\newcommand{\UP}{\mathbf{UP}}
\newcommand{\coUP}{\mathbf{coUP}}

\renewcommand{\P}{\mathbf{P}}
\newcommand{\cep}{\mathbf{C_=P}}
\newcommand{\PSPACE}{\mathbf{PSPACE}}
\newcommand{\BPP}{\mathbf{BPP}}
\newcommand{\BQP}{\mathbf{BQP}}
\newcommand{\EQP}{\mathbf{EQP}}
\newcommand{\QP}{\mathbf{QP}}
\newcommand{\PP}{\mathbf{PP}}
\newcommand{\AWPP}{\mathbf{AWPP}}
\newcommand{\LWPP}{\mathbf{LWPP}}
\newcommand{\GapP}{\mathbf{GapP}}

\newcommand{\intersect}{\cap}
\newcommand{\dom}[1]{\mathop{\mathrm dom}(#1)}

\def\qed{\quad\vrule height8pt width4pt depth1.5pt \bigbreak}

\begin{document}
%
%
\date{}
%
%
\title{\Large\bf Complexity Limitations on Quantum Computation}
\author{\begin{tabular}[t]{c@{\extracolsep{8em}}c} 
Lance Fortnow\thanks{Email: fortnow@cs.uchicago.edu. URL:
    http://www.cs.uchicago.edu/\~{ }fortnow.  Supported
    in part by NSF grant CCR 92-53582. Some of this research occurred
    while the author was visiting the CWI in Amsterdam.}  
& 
John Rogers\thanks{Email: rogers@cs.depaul.edu. URL: 
    http://www.depaul.edu/\~{ }jrogers.} \\
 \\
        Department of Computer Science & School of CTI \\
        University of Chicago & DePaul University \\
        Chicago, IL~~60637 & Chicago, IL~~60604
\end{tabular}}
\maketitle
%
%
\thispagestyle{empty}
\subsection*{\centering Abstract}
{\em
  
  We use the powerful tools of counting complexity and generic oracles
  to help understand the limitations of the complexity of quantum
  computation.  We show several results for the probabilistic
  quantum class $\BQP$.
\begin{itemize}
\item $\BQP$ is low for $\PP$, i.e., $\PP^\BQP=\PP$.
\item There exists a relativized world where $\P=\BQP$ and the
  polynomial-time hierarchy is infinite.
\item There exists a relativized world where $\BQP$ does not have
  complete sets.
\item There exists a relativized world where $\P=\BQP$ but $\P\neq
  \UP\cap\coUP$ and one-way functions exist. This gives a relativized 
  answer to an open question of Simon.
\end{itemize}
}
\section{Introduction}

We have seen a surge in interest in quantum computation over the past
few years. This interest comes from new and good theoretical models
for quantum Turing machines~\cite{BV} and surprising algorithms for
factoring~\cite{Shor} and searching~\cite{Grover}.

Exactly how much can we accomplish with quantum computers? We bring
two powerful tools from computational complexity theory to bear on
this question. First we use counting complexity, in particular the
$\GapP$ functions developed by Fenner, Fortnow and Kurtz~\cite{FFK},
to give us new upper bounds on quantum complexity classes. Next we use
generic oracles to show that the existence of one-way functions does
not necessarily imply that quantum computers are more powerful than
deterministic ones.

The power of a quantum Turing machine lies in its ability to have its
superpositions ``cancel'' each other.  Fenner, Fortnow and
Kurtz~\cite{FFK} developed the notion of $\GapP$ functions, the
closure of the $\#\P$ functions under subtraction. The $\GapP$
functions have some powerful applications based on a similar
cancellation effect.  We show, perhaps not too surprisingly, a close
relationship between $\GapP$-definable counting classes and quantum
computing. We can use this relationship to obtain new limitations on
the complexity of quantum computing.

The usual notion of efficient computation is captured by the bounded
probabilistic polynomial-time Turing machine.  Such a machine accepts
an input $x$ either with probability greater than or equal to $2/3$ or
with probability less than or equal to $1/3$.  In the first case, we
say that $x$ is in the language accepted by $M$ and in the second that
it is not.  The class of languages accepted by these machines is
called $\BPP$.  Replacing the Turing machine with a quantum Turing
machine yields the class $\BQP$.

We show that $\BQP$ is contained in the counting class $\AWPP$.  Based
on previous results about the class $\AWPP$~\cite{Li-PhD}, we can show
that $\BQP$ is low for $\PP$ and so improve the upper bound given by
Adleman, DeMarrais and Huang~\cite{ADH}. We can also use oracle
results about $\AWPP$ to get a relativized world where $\P=\BQP$ but
the polynomial-time hierarchy is infinite. We also use these
techniques to
give a relativized world where $\BQP$ does not have complete sets.

We know that $\BPP \subseteq \BQP$. An important open question is
whether or not the containment is strict.  Showing the containment
strict would require separating $\BPP$ and $\PSPACE$, a presumably
difficult task.  One approach is to investigate what kinds of
conditions would cause a separation between $\BPP$ and $\BQP$.  For
example, Simon~\cite{Simon-quantum} asked whether the existence of
one-way functions is sufficient to cause a separation.  A one-way
function is a one-to-one, honest, polynomial-time computable function
whose inverse is not polynomial-time computable.

Our result shows that there is a relativized world in which this
implication fails.  Although this does not directly refute the
implication, it does demonstrate that, if the implication is true,
proving it will require techniques that do not relativize.

\section{Definitions}

\subsection{Preliminaries}

As usual, $\Sigma$ denotes the alphabet $\{0,1\}$ and $\Sigma^{*}$ the
set of all finite length strings over $\Sigma$. A \emph{language} is a
subset of $\Sigma^{*}$. The notation $|x|$ denotes the length of string
$x$. We will sometimes need to compare strings and natural numbers. To
do so, we will use the polynomial-time computable isomorphism between
the nonzero natural numbers and strings that maps a string $x$ to the
natural number whose binary representation is 1 followed by $x$. 

The notation $\langle m,n \rangle$ denotes the Rogers \cite{Rogers}
pairing function, that is, a polynomial-time computable function that
maps the pair of natural numbers $m$ and $n$ one-to-one and onto the
natural numbers. Note that, given $\langle m,n \rangle$, we can extract
both $m$ and $n$ in polynomial time. Employing the isomorphism defined
above allows us to to apply this pairing function to strings: $\langle
x,y \rangle$ denotes a pairing of strings from which we can easily
extract $x$ and $y$.

\subsection{Models of computation}

Our models of computation are the (classical) Turing machine and the
quantum Turing machine.  Unless otherwise stated, we assume that all
machines are polynomial-time bounded.  See Hopcroft and Ullman
\cite{HoUl} for definitions regarding classical Turing machines.  

We allow the machines to have oracle access, which means that they are
allowed to make membership queries to a language $A$, called the oracle,
and to receive a correct response in constant time. Such machines have a
separate query tape and three extra states: a \emph{query} state, a
\emph{yes} state, and a \emph{no} state. When a machine is computing
relative to an oracle $A$, it can query whether a string $x$ is in $A$
by writing $x$ on the query tape and entering the query state. If $x \in
A$, the computation's next state is the \emph{yes} state, otherwise it
is the \emph{no} state.

A language $L$ is in the complexity class $\BPP$ if there is a
classical machine $M$ such that, for every $x \in L$, at least $2/3$
of the paths in the computation $M(x)$ are accepting paths and, for
every $x \notin L$, no more than $1/3$ of the paths are accepting
paths.

\subsection{Quantum Computation}
\label{quantumsec}

We will use a simplified model of quantum computation due to Bernstein
and Vazirani~\cite{BV}. While simple, this model captures all of the
power of quantum computation. See the paper by Bernstein and
Vazirani~\cite{BV} for a discussion of this model and the physics
underlying it.  For a general introduction to quantum computation see
the survey by Berthiaume~\cite{Berthiaume}.

Consider the transition function of a Turing machine that maps current
state and contents under the tape heads to a new state, new values to
write under the tape heads and a direction to move the heads. A
deterministic Turing machine's transition function has a single
output. A probabilistic Turing machine's transition function maps to a
distribution on outputs with nonnegative probabilities that add up to
one.

A quantum Turing machine's transition function maps to a
\emph{superposition} of the outputs where each output gets an
\emph{amplitude} which may be a complex value.  In the case of $\BQP$
as defined below, Adleman, DeMarrais and Huang~\cite{ADH} and Solovay
and Yao~\cite{SoYa} show that we can assume these amplitudes take one
of the values in
$\{-1,-\frac{4}{5},-\frac{3}{5},0,\frac{3}{5},\frac{4}{5},1\}$.
Bennett, Bernstein, Brassard and Vazirani~\cite{BBBV} show that we can
assume the quantum Turing machine has a single accepting
configuration.

The quantum Turing machines we consider here all run in polynomial
time and thus have an exponential number of possible configurations.

Suppose that before a transition each configuration $C_i$ has a real
amplitude $\alpha_i$. Consider the $L_2$ norm of the amplitudes
\[\sqrt{\sum_{i}\alpha_i^2}\]
A quantum Turing machine is required to preserve this $L_2$ norm.
This is equivalent to the transition matrix $U$ of the configurations
being unitary. For real $U$, $U$ is unitary if the transpose of $U$ is
the inverse of $U$.

To compute the probability of acceptance consider an initial
configuration amplitude vector $\vec{\alpha}$ where $\alpha_0=1$ for
the initial configuration $C_0$ and $\alpha_i=0$ for every other
configuration.  Let $\vec{\beta}=U^t\cdot\vec{\alpha}$ where $t$ is
the running time of the Turing machine. The probability of acceptance
is $\beta_i^2$ where $C_i$ is the accepting configuration.

We can now define $\BQP$ similar to the definition of $\BPP$.
\begin{definition}
A language $L$ is in $\BQP$ if there is a quantum Turing machine $M$
such that for all $x$ in $\Sigma^*$,
\begin{itemize}
\item If $x$ is in $L$ then $M(x)$ accepts with probability at least
  two-thirds.
\item If $x$ is not in $L$ then $M(x)$ accepts with probability at
  most one-third.
\end{itemize}
\end{definition}
Similar to $\BPP$, though with nontrivial proofs, we can assume the
error is exponentially small and that $\BQP$ machines can simulate
deterministic Turing machines and other $\BQP$ machines as subroutines
(see~\cite{BV}).

The class $\EQP$ (sometimes called $\QP$) has the same definition as $\BQP$ 
except that we require zero error.  It is analogous to $\P$ in that every 
computation path halts in polynomial time.  Surprisingly, this class 
appears to be stronger than deterministic polynomial time (see~\cite{BV}).

\subsection{Counting Classes}

A function $f$ from strings to the natural numbers is in the counting
class $\#\P$ if there is a nondeterministic polynomial-time machine $M$
such that $f(x) = m$ iff the computation $M(x)$ has $m$ accepting paths.
In order to understand better the complexity of counting classes like
$\#\P$, Fenner, Fortnow and Kurtz~\cite{FFK} defined the $\GapP$
functions consisting of the closure under subtraction of the set of
$\#\P$ functions (and so a $\GapP$ function's domain is the integers).
Equivalently, $\GapP$ consists of functions $f(x)$ such that for some
nondeterministic polynomial-time Turing machine $M$, $f(x)$ is the
difference between the number of accepting and the number of rejecting
paths of $M(x)$.

The power of $\GapP$ functions lie in their closure properties:
$\GapP$ functions are closed under negation, subtraction, exponential
addition and polynomial multiplication.
\begin{theorem}[Fenner-Fortnow-Kurtz]
\label{gapclosethm}
Let $f$ be a $\GapP$ function and $q$ a polynomial.  Then the
following are $\GapP$ functions: 
\begin{enumerate}
\item $-f(x)$,
\item $\sum_{|y|\leq q(|x|)} f(\langle x,y \rangle)$, and 
\item $\prod_{0\leq y\leq q(|x|)} f(\langle x,y \rangle)$.
\end{enumerate}
\end{theorem}

For the rest of the paper, we will assume that the pairing function is
implicitly used whenever we have a function of two or more arguments.

We can also define many interesting counting classes using $\GapP$
functions. For this paper we consider the following classes.

\begin{definition}
\label{ppdef}
The class $\PP$ consists of those languages $L$ such that for some $\GapP$
function $f$ and all $x$ in $\Sigma^*$,
\begin{itemize}
\item If $x$ is in $L$ then $f(x)>0$.
\item If $x$ is not in $L$ then $f(x)<0$.
\end{itemize}
\end{definition}
The class $\PP$ was first defined by Gill~\cite{Gi} as probabilistic
polynomial time with unbounded error. Definition~\ref{ppdef}, first
given by Fenner, Fortnow, and Kurtz~\cite{FFK}, makes the class
considerably easier to work with.

\begin{definition}
\label{lwppdef}
The class $\LWPP$ consists of those languages $L$ such that for some 
$\GapP$ function $f$, and some polynomial-time computable positive function 
$g$, and for all $x$ in $\Sigma^*$:
\begin{itemize}
\item If $x$ is in $L$ then $f(x)=g(1^{|x|})$.
\item If $x$ is not in $L$ then $f(x)=0$.
\end{itemize}
\end{definition}

\begin{definition}
\label{awppdef}
The class $\AWPP$ consists of those languages $L$ such that for all
polynomials $q$, there is a $\GapP$ function $f$ and a polynomial-time
computable function $g$ such that for all strings $x$ in $\Sigma^*$
and $m\geq |x|$, $0<f(x,1^m)<g(1^m)$ and
\begin{itemize}
\item If $x$ is in $L$ then $f(x,1^m)\geq (1-2^{-q(m)})g(1^m)$.
\item If $x$ is not in $L$ then $f(x,1^m)\leq 2^{-q(m)}g(1^m)$.
\end{itemize}
\end{definition}

The classes $\LWPP$ and $\AWPP$ were first defined by Fenner, Fortnow
and Kurtz~\cite{FFK} and Fenner, Fortnow, Kurtz and Li~\cite{FFKL}.
Though artificial, these classes have some nice properties that we
will use to help classify quantum complexity.

\subsection{One-way functions}

A language $L$ is in the class $\UP$ if there is a classical machine $M$
that, for every $x \in L$, has exactly one accepting path and has no
accepting paths if $x \notin L$.

A polynomial-time computable function $f$ from strings to strings is
\emph{one-way} if it is one-to-one, honest, and not invertible in
polynomial time. Being \emph{honest} means that there is a polynomial
$p$ such that $p(|f(x)|) > |x|$; in other words, honest functions do not
map long input strings to short output strings. Grollmann and Selman
\cite{GrSl} showed that one-way functions exist if and only if $\P \ne
\UP$. Note that these one-way functions may not be suitable for
cryptographic purposes which require average-case hardness against
nonuniform inverters.

\subsection{Generic oracles}

In trying to show that there is an oracle relative to which a particular 
proposition $P$ holds, we often begin by defining an infinite set of 
\emph{requirements}, which are statements about relativized computations.  
An oracle $X$ \emph{satisfies} (or \emph{forces}) a requirement if the 
statement of the requirement is true when the computations are performed 
relative to $X$.  We define the requirements so that, if each is satisfied, 
the proposition $P$ is true.  For example, to make ${\P}^X \ne {\NP}^X$, we 
specify an enumeration of all polynomial-time bounded, deterministic oracle 
Turing machines: $\{M_i\}_{i\in\omega}$.  We then define a nondeterministic 
machine $N$ and an infinite set of requirements $R = \{R_i\}_{i\in\omega}$, 
where $R_i$ is the statement: ``$L(M_i^X) \ne L(N^X)$.''  If we construct 
an oracle $X$ satisfying every $R_i$ then ${\P}^X \ne {\NP}^X$.

Defining the requirements is often quite straightforward.  The
difficulties usually arise when trying to construct the oracle.  We
avoid some of the difficulties by employing \emph{generic} oracles.

A \emph{condition} is a partial function from $\Sigma^{*}$ to
$\{0,1\}$.  A condition $\sigma$ \emph{extends} another condition
$\tau$ if, for all $x \in \dom{\tau}$, $\sigma(x) = \tau(x)$.  An
oracle $A$ extends a condition $\sigma$ if $A$'s characteristic
function extends $\sigma$.  Two conditions $\sigma$ and $\tau$ are
\emph{compatible} if, for all $x \in \dom{\sigma} \intersect
\dom{\tau}$, $\sigma(x) = \tau(x)$.  They \emph{conflict} otherwise.
We will always assume that if a condition is defined on any string of
some length then it is defined on all strings of that length.

A condition $\sigma$ \emph{satisfies} a requirement if any oracle
extending $\sigma$ satisfies it.

A set of conditions $S$ is \emph{dense} if, for every condition
$\tau$, there is a condition $\sigma \in S$ that extends $\tau$.  It
is \emph{definable} if the set $\{\overline\sigma: \sigma \in S\}$
belongs to $\Pi_1^1$ (see \cite{Rogers}).

Restrictions can be set on conditions to achieve a desired separation.
In this paper, we impose the restriction that all conditions have finite
domains. In section \ref{section:one-way}, we will employ
\emph{$\UP\cap\coUP$-conditions}, which have the following further
restrictions: a condition takes on the value 0 for every string not at
an \emph{acceptable length} and it takes on the value 1 for exactly one
string at each acceptable length. An acceptable length is an integer in
the range of the $tower$ function, which has the recursive definition:
$tower(0)=2$, $tower(n+1) = 2^{tower(n)}$. That is, $tower(n)$ is a
tower of 2's with height $n+1$. 

An oracle $A$ \emph{meets} a set of conditions $S$ if there is some
$\sigma$ in $S$ that is extended by $A$. A \emph{generic} oracle is one
that meets every dense definable set of conditions. A
\emph{$\UP\cap\coUP$-generic} oracle is one that meets every dense
definable set of $\UP\cap\coUP$-conditions. $\UP\cap\coUP$-generics were
first developed by Fortnow and Rogers~\cite{FoRo} to study the
relationship between separability and one-way functions. More background
about these oracles and a variety of other generic oracles can be found
in that earlier paper and in papers by Blum and Impagliazzo \cite{BI}
and Fenner, Fortnow, and Kurtz \cite{FFK}. 

\section{Counting Complexity}

In this section we show a close connection between counting complexity
and quantum computing.

\begin{theorem}
\label{awppthm}
\ \[\BQP\subseteq\AWPP\]
\end{theorem}

Theorem~\ref{awppthm} follows from the following lemma.

\begin{lemma}
\label{gaplem}
For any quantum Turing machine $M$ running
in time bounded by a polynomial $t(n)$, there is a $\GapP$
function $f$ such that for all inputs $x$,
\[\Pr(M(x)\mbox{ accepts})=\frac{f(x)}{5^{2t(|x|)}}.\]
\end{lemma}

Proof of Theorem~\ref{awppthm}. 
Fix a language $L$ in $\BQP$ and a polynomial $q$. Let $M$ be a
polynomial-time quantum Turing machine that on input $(x,1^m)$ accepts
for $x$ in $L$ with probability at least $1-2^{-q(m)}$ and accepts for
$x$ not in $L$ with probability at most $2^{-q(m)}$.

Fix $x$ and $m$ with $m\geq |x|$. Then there is a polynomial $t(m)$ that
bounds the running time of $M(x,1^m)$. By Lemma~\ref{gaplem} there is a
$\GapP$ function $f$ such that $f(x,1^m)/5^{2t(m)}$ is the acceptance
probability of $M(x,1^m)$. We can thus fulfill the requirements of
Definition~\ref{awppdef} by letting $g(1^m)=5^{2t(m)}$. \qed

Proof of Lemma~\ref{gaplem}.
We can assume that $M$ has at most $2^t$ configurations.  Let
$U$ be the transition matrix of $M$. By the discussion in
Section~\ref{quantumsec} we can assume the entries of $U$ are of the
form $w/5$ for $w$ an integer between $-5$ and $5$.  By the nature of
a transition matrix, we can compute the $(i,j)$ entry of $U$ in time
polynomial in $|x|$.

Let $V=5U$ so $V$ has only integral entries. Let $\vec{\alpha}$ be the
initial configuration amplitude vector as described in
Section~\ref{quantumsec}. Let $\vec{\beta}=V^t\cdot\vec{\alpha}$. Note
that each $\beta_i$, a component of $\vec{\beta}$ corresponding to
configuration $C_i$, is an exponential sum of a polynomial product of
polynomial-time computable entries of $V$. By
Theorem~\ref{gapclosethm}, we have that each $\beta_i$ is a $\GapP$
function.

Let $f(x)$ be $\beta^2_i$ where $C_i$ is the accepting configuration of
$M(x)$. Again by Theorem~\ref{gapclosethm} we have $f(x)$ is a $\GapP$
function. We have that $f(x,m)/5^{t(|x|)^2}$ is the acceptance
probability of $M(x)$. \qed

We can now use properties of $\AWPP$ to better understand the
complexity of $\BQP$. Lide Li~\cite{Li-PhD} gave an upper bound on the
complexity of $\AWPP$.

\begin{theorem}[Li]
\label{lithm}
$\AWPP$ is low for $\PP$, i.e., $\PP^\AWPP=\PP$.
\end{theorem}

For completeness we sketch the proof of Theorem~\ref{lithm}.

Proof Sketch. Suppose $L$ is in $\PP^A$ for some $A$ in $\AWPP$. By
Definition~\ref{ppdef}, there is some $h\in\GapP^A$ such that for
$x\in L$, $h(x)\geq 1$ and $h(x)\leq -1$ otherwise. Let $M^A$ be a
relativized nondeterministic polynomial-time Turing machine such that
$h(x)$ is the difference of the number of accepting and rejecting
computations of $M^A(x)$. We assume without loss of generality that
for every $A$ and $x$ each computation path of $M^A(x)$ makes the same
number of queries.

Pick a polynomial $q(n)$ such that for strings of length $n$, $M^A$
has less than $2^{q(n)/2}$ computation paths. Let $f$ and $g$ be
$\GapP$ and polynomial-time computable functions defined for $A$ and
$q$ as in Definition~\ref{awppdef}.  Let $N$ be a nondeterministic
polynomial-time Turing machine such that $f(x,1^m)$ is the difference
of the number of accepting and rejecting paths of $N(x,1^m)$.

Create a new nondeterministic polynomial-time Turing machine $M'$ as
follows. On input $x$, simulate $M^A(x)$. Every time $M$ makes a query
$y$ to $A$, simulate $N(y,1^{|x|})$. If $N$ accepts then continue the
computation of $M$ assuming $y$ is in $A$. If $N$ rejects then
continue the computation of $M$ assuming $y$ is not in $A$.


By the choice of $q$, the mistakes made by the wrong simulation, even
totaled over every computation path of $M^A(x)$, are not enough to
affect the sign of the difference of the number of accepting and
rejecting paths of $M'$. \qed

{From} Theorem~\ref{lithm} we get the same result for $\BQP$.
\begin{corollary}
$\BQP$ is low for $\PP$.
\end{corollary}

This improves and simplifies the bound given by Adleman, DeMarrais and
Huang~\cite{ADH}.

\begin{corollary}[Adleman-DeMarrais-Huang]
\ \[\BQP\subseteq\PP\subseteq\P^{\#\P}\subseteq\PSPACE\]
\end{corollary}

We also have a class containing $\BQP$ that is not known to contain
$\NP$ as Beigel~\cite{Bei} has a relativized world where $\NP$ is not
low for $\PP$.

Fenner, Fortnow, Kurtz and Li~\cite{FFKL} give an interesting collapse
for $\AWPP$ relative to generic oracles. Their proof builds on a
connection between decision tree complexity and low-degree polynomials
developed by Nisan and Szegedy~\cite{NS}.

\begin{theorem}[FFKL,NS]
\label{awppgenthm}
If $\P=\PSPACE$ (unrelativized) then $\P^G=\AWPP^G$ for any generic
$G$.
\end{theorem}

We can create an oracle $H$ by starting with an oracle making
$\P=\PSPACE$ and joining a generic $G$ to that. Because the
polynomial-time hierarchy is infinite relative to generic oracles and
because Theorem~\ref{awppthm} relativizes, we can get some interesting
relativized worlds.

\begin{corollary}
There exists a relativized world where $\P=\BQP$ and the
polynomial-time hierarchy is infinite.
\end{corollary}

This greatly strengthens the result of Bennett, Bernstein, Brassard and
Vazirani~\cite{BBBV} giving a relativized world where $\NP$ is not in
$\BQP$.

Using a proof similar to that of Theorem~\ref{awppthm} we get a
stronger upper bound for $\EQP$.
\begin{theorem}
\label{lwppthm}
\ \[\EQP\subseteq\LWPP\]
\end{theorem}

Whether Graph Isomorphism can be solved quickly by quantum computers
remains an interesting open question. This possibility is consistent
with Theorem~\ref{lwppthm} as K\"{o}bler, Sch\"{o}ning and
Tor\'{a}n~\cite{KST} have show that Graph Isomorphism sits in $\LWPP$.

\subsection{Extensions}

The techniques in our paper can also be used to show bounds on the
decision tree complexity of quantum computers. Here we consider the
situation where we wish to compute a function $f:\{
0,1\}^N\rightarrow\{ 0,1\}$ where access to input bits are only via
oracle questions. We typically do not care about running time in this
model, only the maximum number of queries on any computation path.

Grover~\cite{Grover} shows how to get a nontrivial advantage with
quantum computers: He shows that computing the OR function needs only
$O(\sqrt{N})$ queries although deterministically all $N$ input bits
are needed in the worst case.

Bernstein and Vazirani~\cite{BV} give a superpolynomial gap and
Simon~\cite{Simon-quantum} gives an exponential gap. However, both of
these gaps require that there are particular subsets of the inputs to
which $f$ is restricted.

Beals, Buhrman, Cleve, Mosca and de Wolf~\cite{BBCMW} notice that a
limitation on the decision tree complexity of quantum computation
follows from the techniques of the proof of Theorem~\ref{awppgenthm}.
\begin{corollary}
\label{dtcor}
If there is a quantum algorithm computing a function $f$ defined on all
of $\{0,1\}^N$ and using $t$ queries then there exists a deterministic
algorithm computing $f$ using $O(t^8)$ queries.
\end{corollary}

Using other techniques, Beals, Buhrman, Cleve, Mosca and de
Wolf~\cite{BBCMW} improve Corollary~\ref{dtcor} to $O(t^6)$ queries and
show better bounds for specific functions.

Vereshchagin~\cite{Ver-complete} gives the following useful lemma for
proving a relativized lack of complete sets for some classes.
\begin{lemma}[Vereshchagin]
\label{verlemma}
Under some weak restrictions on complexity classes ${\cal A}$ and
${\cal B}$, if
\begin{itemize}
\item the class of boolean functions computed by
polylogarithmic depth decision trees of type ${\cal A}$
coincides with the class of functions computed by
deterministic polylogarithmic depth decision trees and
\item there exists a promise problem
solved by polylogarithmic depth decision trees of
of type ${\cal B}$ but not by
deterministic polylogarithmic depth decision trees
\end{itemize}
then there is an oracle where ${\cal A}$ does not have sets
polynomial-time Turing hard for ${\cal B}$.

\end{lemma}

The following result then follows from Corollary~\ref{dtcor} and
Lemma~\ref{verlemma}.
\begin{corollary}
There exists a relativized world where $\BQP$ has no hards sets for
$\BPP$. In particular, $\BQP$ has no complete sets in this world.
\end{corollary}

Lemma~\ref{gaplem} shows how to compute the probability acceptance of
a quantum Turing machine with a $\GapP$ function. Fenner, Green, Homer
and Pruim~\cite{FGHP} give a result in the other direction.

\begin{theorem}[FGHP]
\label{fghpthm}
For any $\GapP$ function $f$ there exists a
  polynomial-time quantum Turing machine $M$ and a polynomial $p$ such
  that for all $x$,
\[\Pr(M(x)\mbox{ accepts})=\frac{f(x)}{2^{p(|x|)}}.\]
\end{theorem}
Theorem~\ref{fghpthm} creates a quantum machine with amplitudes
contained in $\{0, -1, -\frac{1}{\sqrt{2}}, \frac{1}{\sqrt{2}}, 1\}$.
Fenner, Green, Homer and Pruim~\cite{FGHP} note that our
Lemma~\ref{gaplem} holds where amplitudes may be any
positive or negative square roots of rational numbers (the value ``5''
in the statement of Lemma~\ref{gaplem} may have to be replaced with a
different positive integer).

From Lemma~\ref{gaplem} and Theorem~\ref{fghpthm} we immediately get a
new characterization of the class $\cep$. The class $\cep$ consists of
the languages $L$ for which there exists a $\GapP$ function $f$ such
that $x$ is in $L$ exactly when $f(x)=0$.
\begin{corollary}
A language $L$ is in $\cep$ if and only if
there exists a polynomial-time quantum Turing machine $M$ such that $x$
is in $L$ exactly when the probability that $M(x)$ accepts is zero.
\end{corollary}

Watrous~\cite{Watrous} proves similar results for space-bounded quantum
Turing machines.

\section{One-Way Functions}
\label{section:one-way}

We show that one-way functions are not sufficient to guarantee the
hardness of $\BQP$.
\begin{theorem}
\label{theorem:main}
There is an oracle $C$ relative to which one-way functions exist and
$\P^C = \BQP^C$.
\end{theorem}

Thus, to demonstrate that the existence of one-way functions implies a
separation between $\BPP$ and $\BQP$ will require nonrelativizing
techniques.  

We actually prove a stronger result from which
Theorem~\ref{theorem:main} follows.
\begin{theorem}
\label{upthm}
There is an oracle $C$ relative to which
$\P^C=\BPP^C=\BQP^C\neq\UP^C\cap\coUP^C$.
\end{theorem}

To prove Theorem \ref{upthm}, we need the following theorem
due to Bennett, Bernstein, Brassard and Vazirani~\cite{BBBV}.
\begin{theorem}[BBBV]
\label{theorem:BBBV}
Let $M$ be an oracle $\BQP$ machine that runs in time $p(n)$ and let $O$
be an oracle and $x$ an $n$-bit input. There is a set $S$ such that for
all oracles $O'$, if $O'$ differs from $O$ only on a single string and
that string is not in $S$ then $|P[M^{O'}$ accepts $x] - P[M^O$ accepts
$x]| \le \epsilon$, where $|S| \le 4p^2(n)/\epsilon^{2}$.
\end{theorem}

This theorem states that for an oracle $\BQP$ Turing machine $M$ and
an input $x$ whose length is $n$, there is a polynomial (in $n$) sized
set $S$ such that, if a string $y$ is not in $S$, we can change the
oracle's answer on $y$ and the probability that $M$ accepts $x$ is
still bounded away from $1/2$.

Proof of Theorem~\ref{upthm}.  Let $H$ be an oracle relative to which
$\P = \PSPACE$ ($H$ can be any $\PSPACE$-complete language).  Let $G$
be a $\UP\cap\coUP$-generic oracle, which must have exactly one string
at lengths that are exponentially far apart.  Let $C = H \oplus G =
\{0x|x\in H\} \cup \{1y|y\in G\}$.  The oracle $C$ represents a
relativization that identifies $\P$ and $\PSPACE$ (and so $\P = \BPP =
\BQP$) and a \emph{re-relativization} that, we will show, separates
$\P$ and $\UP\cap\coUP$ but that still leaves $\P = \BPP = \BQP$.

First we show that $\P^C \ne \UP^C\cap\coUP^C$. Let $L^X =
\{0^n|(\exists x)|x| = n-1 \;\&\; x0 \in X\}$. It's easy to see that
$L^G \in \UP^G\cap\coUP^G$ and so is in $\UP^C\cap\coUP^C$. A simple
diagonalization argument demonstrates that $L^G \notin \P^G$. Because
$G$ is generic with respect to $H$, $L^C \notin \P^C$.

Next we show that $\P^C = \BQP^C$.  Let $M$ be a $\BQP^C$ machine that
runs in time $p(n)$.  Since $G$ is generic we can assume that $M$ is
\emph{categorically} a $\BQP$ machine, i.e., for any oracle $A$ and
input $x$, $M^A(x)$ accepts with probability greater than or equal to
$2/3$ or less than or equal to $1/3$ (see~\cite{BI}).

Let $x$ be an input of length $n$.  We need to show that there is a 
deterministic polynomial-time machine $N$ that, relative to $C$, determines 
for an input $x$ whether $M^C(x)$ accepts.  Because $M$ runs in polynomial 
time, there are a polynomial number of lengths for strings that $M$ can 
query in an oracle.  Because $G$ has exactly one string at every acceptable 
length, there are polynomially many strings in $G$ that could affect $M$'s 
computation on $x$.  Because the strings in $G$ are exponentially far 
apart, all but at most one are at lengths that are so short that $N$ on 
input $x$ can query $G$ on every string at those lengths and so find all of 
them.

So the only string that $N$ needs to worry about is one at a length
$\ell$ that is so large that $N$ would have to query exponentially many
strings to be certain of finding it. Call this string $y$. Even though
$N$ cannot find $y$ by searching, it \emph{can} use its access to $H$ to
figure out what $M$ would do on input $x$ under the assumption that
there are no strings of length $\ell$ in $G$.

Let us say that, under this assumption, $M(x)$ accepts.  However, we know 
there is a string of length $\ell$ in $G$ that could cause $M$ to change 
its computation and reject $x$.  But Theorem~\ref{theorem:BBBV} says that 
there is a set of strings $S$ whose cardinality is bounded by 
$4p^2(n)/\epsilon^{2}$ such that, if $y$ is not in $S$, the probability 
that $M$ changes its computation is less than or equal to $\epsilon$.

Set $\epsilon$ to a value strictly less than $1/6$ and say that we know 
that $y$ (if it exists) is not in $S$.  By Theorem~\ref{theorem:BBBV}, the 
probability that $M$ accepts $x$ is still strictly greater than $1/2$.  In 
other words, $x$ is still in the language accepted $M$ relative to $G$.  So 
if $N$ knows that $y$ is not in $S$, it can simply run the simulation of 
$M(x)$ under the assumption that there is no string of length $\ell$ in $G$ 
and output the correct answer.

So how does $N$ determine whether $y$ is in $S$?  It asks for an explicit 
enumeration of $S$.  That is, it asks the question: ``What is the set $S$ 
of strings of length $\ell$ such that, if one of those strings is in the 
oracle, $M$ rejects $x$?''  $S$ has size at most $4c^{2}p^{2}(n)$, where 
$c>6$.  This question can be answered in $\PSPACE$ without querying $G$.  
$N$ can use its access to $H$ to find $S$ in polynomial time.  It then 
queries $G$ for each of those strings.  If none of those strings are in $G$ 
then $N$ accepts input $x$ because that is what $M$ would do.  If $N$ finds 
one of those strings in $G$, it would then be able to simulate the 
computation of $M(x)$ with full knowledge of all of the strings in $G$ that 
could possibly affect that computation.  \qed

\subsection{Cryptographic One-Way Functions}

The assumption $\P\neq\UP$ does not necessarily imply the existence of 
{\em cryptographic} one-way functions, i.e., functions not invertible
on a large fraction of inputs with nonuniform polynomial-size
circuits. Whether there exists a relativized world where $\BPP=\BQP$
and cryptographic one-way functions exist remains an interesting open
question.

One possible approach would look at whether $\P=\BQP$ relative to a
random oracle since relative to a random oracle cryptographic one-way
functions exist~(see \cite{IR}). Showing this would imply that
factoring is in $\BQP=\BPP$ and thus factoring is efficiently
computable on probabilistic machines~\cite{Shor}.

\begin{theorem}
If $\P=\BQP$ relative to random oracle then $\BQP=\BPP$
(unrelativized).
\end{theorem}

Proof. Let $L$ be in $\BQP$, then for every oracle $R$, $L$ is in
$\BQP^R$. Thus by assumption $L$ is in $\P^R$ for most oracles
$R$. Bennett and Gill~\cite{BeGi} show that every language with this
property sits in $\BPP$. \qed

However, we could possibly prove $\P=\BQP$ for random oracles under
some assumption like $\P=\PSPACE$. If this were true with a
relativizable proof, we could start with an oracle relativize to which
$\P=\PSPACE$ and join a random oracle to it. This would yield a
relativized world where $\P=\BQP$ and cryptographic one-way functions
exist.

\section{Conclusions}
We give results in this paper indicating severe restrictions on the
complexity of quantum computing. We conjecture that $\BQP$ actually
contains \emph{no} interesting complexity classes outside of $\BPP$.

Still we believe that quantum computing remains a potentially powerful
model of computation. Quantum computers can quickly solve some
problems not known complete such as factoring~\cite{Shor} and the
potential to solve problems such as graph isomorphism and finding a
short vector in a lattice. Also quantum computing can give a large
increase in speed, for example a quadratic improvement in $\NP$-like
search problems~\cite{Grover}.

\section*{Acknowledgments}

We would like to thank Andr\'e Berthiaume, Harry Buhrman, Richard
Cleve, Ronald de Wolf, Wim van Dam and John Watrous for a number of
illuminating conversations on quantum computation.  Also, thanks to
Ronald de Wolf for the corrected statement of
Theorem~\ref{theorem:BBBV}.


\begin{thebibliography}{BBC{\etalchar{+}}98}

\bibitem[ADH97]{ADH}
L.~Adleman, J.~{DeMarrais}, and M.~Huang.
\newblock Quantum computability.
\newblock {\em SIAM Journal on Computing}, 26(5):1524--1540, 1997.

\bibitem[BBBV97]{BBBV}
C.~Bennett, E.~Bernstein, G.~Brassard, and U.~Vazirani.
\newblock Strengths and weaknesses of quantum computing.
\newblock {\em SIAM Journal on Computing}, 26(5):1510--1523, 1997.

\bibitem[BBC{\etalchar{+}}98]{BBCMW}
R.~Beals, H.~Buhrman, R.~Cleve, M.~Mosca, and R.~de~Wolf.
\newblock Quantum lower bounds by polynomials.
\newblock In {\em Proceedings of the 39th IEEE Symposium on Foundations of
  Computer Science}, New York, 1998. IEEE.
\newblock To appear.

\bibitem[Bei94]{Bei}
R.~Beigel.
\newblock Perceptrons, {PP} and the polynomial hierarchy.
\newblock {\em Computational Complexity}, 4:314--324, 1994.

\bibitem[Ber97]{Berthiaume}
A.~Berthiaume.
\newblock Quantum computation.
\newblock In A.~Selman and L.~Hemaspaandra, editors, {\em Complexity Theory
  Retrospective II}, pages 23--51. Springer, 1997.

\bibitem[BG81]{BeGi}
C.~Bennett and J.~Gill.
\newblock {Relative to a random oracle, $P^A\not=NP^A\not=co-NP^A$ with
  probability one}.
\newblock {\em SIAM Journal on Computing}, 10:96--113, 1981.

\bibitem[BI87]{BI}
M.~Blum and R.~Impagliazzo.
\newblock Generic oracles and oracle classes.
\newblock In {\em Proceedings of the 28th IEEE Symposium on Foundations of
  Computer Science}, pages 118--126. IEEE, New York, 1987.

\bibitem[BV97]{BV}
E.~Bernstein and U.~Vazirani.
\newblock Quantum complexity theory.
\newblock {\em SIAM Journal on Computing}, 26(5):1411--1473, 1997.

\bibitem[FFK94]{FFK}
S.~Fenner, L.~Fortnow, and S.~Kurtz.
\newblock Gap-definable counting classes.
\newblock {\em Journal of Computer and System Sciences}, 48(1):116--148, 1994.

\bibitem[FFKL93]{FFKL}
S.~Fenner, L.~Fortnow, S.~Kurtz, and L.~Li.
\newblock An oracle builder's toolkit.
\newblock In {\em Proceedings of the 8th IEEE Structure in Complexity Theory
  Conference}, pages 120--131. IEEE, New York, 1993.

\bibitem[FGHP98]{FGHP}
S.~Fenner, F.~Green, S.~Homer, and R.~Pruim.
\newblock Determining acceptance possibility for a quantum computation is hard
  for {PH}.
\newblock Technical Report 98-008, Computer Science Department, Boston
  University, 1998.

\bibitem[FR94]{FoRo}
L.~Fortnow and J.~Rogers.
\newblock Separability and one-way functions.
\newblock In {\em Proceedings of the 5th Annual International Symposium on
  Algorithms and Computation}, volume 834 of {\em Lecture Notes in Computer
  Science}, pages 396--404. Springer, Berlin, 1994.

\bibitem[Gil77]{Gi}
J.~Gill.
\newblock Computational complexity of probabilistic complexity classes.
\newblock {\em SIAM Journal on Computing}, 6:675--695, 1977.

\bibitem[Gro96]{Grover}
L.~Grover.
\newblock A fast quantum mechanical algorithm for database search.
\newblock In {\em Proceedings of the 28th ACM Symposium on the Theory of
  Computing}, pages 212--219. ACM, New York, 1996.

\bibitem[GS88]{GrSl}
J.~Grollmann and A~Selman.
\newblock Complexity measures for public-key cryptosystems.
\newblock {\em SIAM Journal on Computing}, 17:309--355, 1988.

\bibitem[HU79]{HoUl}
J.~Hopcroft and J.~Ullman.
\newblock {\em Introduction to Automata Theory, Languages and Computation}.
\newblock Addison-Wesley, Reading, Mass., 1979.

\bibitem[IR89]{IR}
R.~Impagliazzo and S.~Rudich.
\newblock Limits on the provable consequences of one-way permutations.
\newblock In {\em Proceedings of the 21st ACM Symposium on the Theory of
  Computing}, pages 44--61. ACM, New York, 1989.

\bibitem[KST92]{KST}
J.~{K\"{o}bler}, U.~{Sch\"{o}ning}, and J.~{Tor\'{a}n}.
\newblock Graph isomorphism is low for {PP}.
\newblock {\em Computational Complexity}, 2(4):301--330, 1992.

\bibitem[Li93]{Li-PhD}
L.~Li.
\newblock {\em On the counting functions}.
\newblock PhD thesis, University of Chicago, 1993.
\newblock Department of Computer Science TR 93-12.

\bibitem[NS94]{NS}
N.~Nisan and M.~Szegedy.
\newblock On the degree of boolean functions as real polynomials.
\newblock {\em Computational Complexity}, 4(4):301--313, 1994.

\bibitem[Rog87]{Rogers}
H.~Rogers.
\newblock {\em Theory of Recursive Functions and Effective Computability}.
\newblock MIT Press, Cambridge, Massachusetts, 1987.

\bibitem[Sho97]{Shor}
P.~Shor.
\newblock Polynomial-time algorithms for prime factorization and discrete
  logarithms on a quantum computer.
\newblock {\em SIAM Journal on Computing}, 26(5):1484--1509, 1997.

\bibitem[Sim97]{Simon-quantum}
D.~Simon.
\newblock On the power of quantum computation.
\newblock {\em SIAM Journal on Computing}, 26(5):1474--1483, 1997.

\bibitem[SY96]{SoYa}
R.~Solovay and A.~Yao, 1996.
\newblock Manuscript.

\bibitem[Ver94]{Ver-complete}
N.~Vereshchagin.
\newblock Relativizable and non-relativizable theorems in the polynomial theory
  of algorithms.
\newblock {\em Russian Academy of Sciences Izvestiya Mathematics},
  42(2):261--298, 1994.

\bibitem[Wat98]{Watrous}
J.~Watrous.
\newblock Relationships between quantum and classical space-bounded complexity
  classes.
\newblock In {\em Proceedings of the 13th IEEE Conference on Computational
  Complexity}, pages 210--227. IEEE, New York, 1998.

\end{thebibliography}

\newcommand{\etalchar}[1]{$^{#1}$}

\end{document}